\documentclass[aps,prl,twocolumn,superscriptaddress,showpacs]{revtex4}

\usepackage{graphicx}

\begin{document}

\title{Dynamical Instability of a Doubly Quantized Vortex in a Bose-Einstein condensate}

\author{Y. Shin}
\affiliation{MIT-Harvard Center for Ultracold Atoms, Research
Laboratory of Electronics, Department of Physics, Massachusetts
Institute of Technology, Cambridge, Massachusetts, 02139}
\author{M. Saba}
\affiliation{MIT-Harvard Center for Ultracold Atoms, Research
Laboratory of Electronics, Department of Physics, Massachusetts
Institute of Technology, Cambridge, Massachusetts, 02139}
\author{M. Vengalattore}
\affiliation{MIT-Harvard Center for Ultracold Atoms, Jefferson
Laboratory, Physics Department, Harvard University, Cambridge,
Massachusetts, 02138}
\author{T.~A. Pasquini}
\affiliation{MIT-Harvard Center for Ultracold Atoms, Research
Laboratory of Electronics, Department of Physics, Massachusetts
Institute of Technology, Cambridge, Massachusetts, 02139}
\author{C. Sanner}
\affiliation{MIT-Harvard Center for Ultracold Atoms, Research
Laboratory of Electronics, Department of Physics, Massachusetts
Institute of Technology, Cambridge, Massachusetts, 02139}
\author{A.~E. Leanhardt}
\affiliation{MIT-Harvard Center for Ultracold Atoms, Research
Laboratory of Electronics, Department of Physics, Massachusetts
Institute of Technology, Cambridge, Massachusetts, 02139}
\author{M. Prentiss}
\affiliation{MIT-Harvard Center for Ultracold Atoms, Jefferson
Laboratory, Physics Department, Harvard University, Cambridge,
Massachusetts, 02138}
\author{D.~E. Pritchard}
\affiliation{MIT-Harvard Center for Ultracold Atoms, Research
Laboratory of Electronics, Department of Physics, Massachusetts
Institute of Technology, Cambridge, Massachusetts, 02139}
\author{W. Ketterle}
\homepage[URL: ]{http://cua.mit.edu/ketterle_group/}
\affiliation{MIT-Harvard Center for Ultracold Atoms, Research
Laboratory of Electronics, Department of Physics, Massachusetts
Institute of Technology, Cambridge, Massachusetts, 02139}

\date{\today}

\begin{abstract}
Doubly quantized vortices were topologically imprinted in
$|F=1\rangle$ $^{23}$Na condensates, and their time evolution was
observed using a tomographic imaging technique. The decay into two
singly quantized vortices was characterized and attributed to
dynamical instability. The time scale of the splitting process was
found to be longer at higher atom density.
\end{abstract}


\maketitle

Quantum fluids, like superfluid He, electrons in a superconductor
or a Bose-Einstein condensate of atoms, are described by a
macroscopic wavefunction. This requires the flow field to be
irrotational, and gives rise to superfluidity and quantized
circulation~\cite{NOP90}. Atoms in a Bose-Einstein condensate, for
example, can only circulate with angular momentum equal to integer
multiple of $\hbar$, in the form of a quantized
vortex~\cite{PES02}.

Vortices are excited states of motion and therefore energetically
unstable towards relaxation into the motional ground state, where
the condensate is at rest. However, quantization constrains the
decay: a vortex in Bose-Einstein condensates cannot simply fade
away or disappear, it is only allowed to move out of the
condensate or annihilate with another vortex of opposite
circulation. Vortex decay and metastability, due to inhibition of
decay, have been a central issue in the study of
superfluidity~\cite{Rok97,DBE97,PLE99,GP99,BR99,VSS01}. In almost
pure Bose-Einstein condensates, vortices with lifetimes up to tens
of seconds have been observed~\cite{MAH99,MCW00,ARV01}.

Giving a Bose-Einstein condensate angular momentum per particle
larger than $\hbar$ can result in one multiply-quantized vortex
with large circulation or, alternatively, in many singly-quantized
vortices each with angular momentum $\hbar$. The kinetic energy of
atoms circulating around the vortex is proportional to the square
of the angular momentum; therefore the kinetic energy associated
with the presence of a multiply-quantized vortex is larger than
the kinetic energy of a collection of singly-quantized vortices
carrying the same angular momentum. A multiply-quantized vortex
can decay coherently by splitting into singly-quantized vortices
and transferring the kinetic energy to coherent excitation modes,
a phenomenon called dynamical instability which is driven by
atomic interactions~\cite{PLE99,SVS02,MMI03,IHS03}, and not caused
by dissipation in an external bath. Observations of arrays of
singly-quantized vortices in rapidly rotating condensates
\cite{MCW00,ARV01} indirectly suggests that the dynamical
instability leads to fast decay of multiply-quantized vortices.
However, the existence of stable multiply-quantized vortices in
trapped Bose-Einstein condensates has been predicted with a
localized pinning potential~\cite{SVS02} or in a quartic
potential~\cite{Lund02}. Stable doubly-quantized vortices were
observed in superconductors in presence of pinning
forces~\cite{BMJ95} and in superfluid $^3$He-A which has a
multicomponent order parameter~\cite{BEK00}. Recently, formation
of a multiply-quantized vortex in a Bose-Einstein condensate has
been demonstrated using topological phases~\cite{INO00,LGC02}, and
surprisingly long lifetime of a ``giant'' vortex core has been
reported~\cite{ECH03}. The study of topological excitation and
their stability is an active frontier in the field of quantum
degenerate gases~\cite{KS01,RA03}.

In this Letter, we study the time evolution of a doubly-quantized
vortex state in a Bose-Einstein condensate, and directly confirm
its dynamical instability by observing that a doubly-quantized
vortex core splits into two singly-quantized vortex cores. The
characteristic time scale of the splitting process was determined
as a function of atom density and was longer at higher atomic
density.

\begin{figure}
\begin{center}
\includegraphics{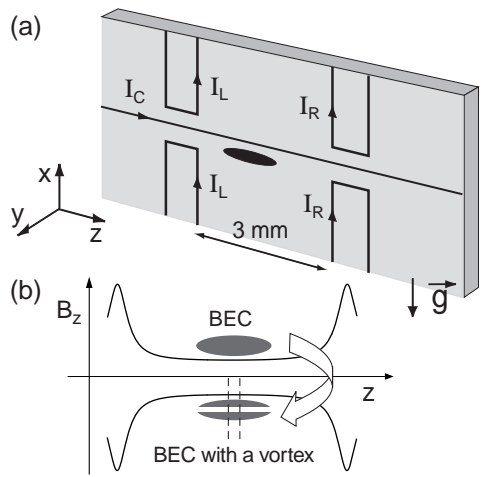}
\caption{(a) Wire pattern on the atom chip. A magnetic trap is
formed by a current $I_C$ flowing through the center wire in
conjunction with a external uniform magnetic field $B_x$. The
axial confinement along $z$ direction is generated by currents
$I_L$ and $I_R$ in the end-cap wires. Each current is controlled
independently. A 2~$\mu$m thick Au film was deposited by
evaporation on a thermally oxidized Si substrate and wires were
patterned by photolithography and wet etching. The width of the
center wire and the end-cap wires were 50~$\mu$m and 100~$\mu$m,
respectively. (b) Imprinting of a vortex in a Bose-Einstein
condensate. By inverting the $z$ direction magnetic field $B_z$, a
doubly quantized vortex was imprinted in $|F=1\rangle$
condensates, using topological phases as in Ref.~\cite{LGC02}. The
direction of $I_{L}$ and $I_{R}$ were also reversed to maintain
the axial confinement. The dashed lines indicate the selective
probing region for tomographic imaging as described in the
text.\label{f:chip}}
\end{center}
\end{figure}

Bose-Einstein condensates containing over $10^7$ $^{23}$Na atoms
were created in the $|F = 1,m_F = -1\rangle$ state, transferred
into an auxiliary chamber~\cite{GCL02}, and loaded into a
Ioffe-Pritchard magnetic trap generated by a microfabricated atom
chip~\cite{OFS01,HHH01,LCK02}. The wire pattern on the atom chip
is shown in Fig.~\ref{f:chip}(a). In our previous work, we used a
Z-shaped wire trap where changing the sign of the axial magnetic
field curvature was technically impossible so that we could not
trap condensates after imprinting a vortex. To overcome this
technical difficulty, we designed our new chip with separate
end-cap wires, allowing independent control of the axial magnetic
field. Typical wire currents were $I_{C}=1.53$~A in the center
wire and $I_{L}=I_{R}=0.1$~A in the end-cap wires, and the
external magnetic field was $B_z=450$~mG and $B_x=5.3$~G,
resulting in a radial (axial) trap frequency $f_{r}=220$~Hz
($f_{z}=3$~Hz) and a distance of the trap from the chip surface
$d=600~\mu$m. After holding condensates for 2~s to damp
excitations which might have been caused by the loading process,
condensates contained over $1.5 \times 10^6$ atoms and the
lifetime of condensates was $\approx 8$~s with a radio-frequency
(rf) shield~\cite{KDS99}.

Doubly-quantized vortices were topologically imprinted in
condensates by inverting the axial magnetic field, $B_z$, as
demonstrated in Ref.~\cite{LGC02}. $B_z$ was ramped linearly from
450~mG to $-460$~mG in 12~ms. As $B_z$ passed zero, the sign of
axial field curvature was changed by reversing the directions of
$I_L$ and $I_R$ in 1~ms. The trap center position and the axial
trap frequency of the inverted wire trap were matched to those of
the original wire trap by adjusting the final values for $I_L$ and
$I_R$. Losses due to nonadiabatic spin flips as $B_z$ passed
through zero reduced the number of atoms in the condensate after
imprinting to about $\sim 1 \times 10^6$, giving a typical healing
length $\xi=0.4~\mu$m. The lifetime of condensates after
imprinting was less than 2~s.

The vortex imprinting process was accompanied by a sudden
mechanical squeeze in the radial direction and a kick in the
vertical direction. The radial trap frequency is proportional to
the square root of the bias magnetic field ($f_r \propto
|B_z|^{-1/2}$) and became temporarily higher during field
inversion. Additionally, the vertical position of the trap center
changed as the gravitational sag ($\propto f_r^{-2}$) changed from
$5.1~\mu$m to zero. The Thomas-Fermi radius of condensates in the
loading phase was $\sim 5~\mu$m. After imprinting a vortex, the
amplitude of quadruple oscillation in the axial direction was
$\sim 20~\%$ of the axial length of condensates ($\approx
600~\mu$m), but there was no detectable dipole oscillation in the
vertical direction.

\begin{figure}
\begin{center}
\includegraphics{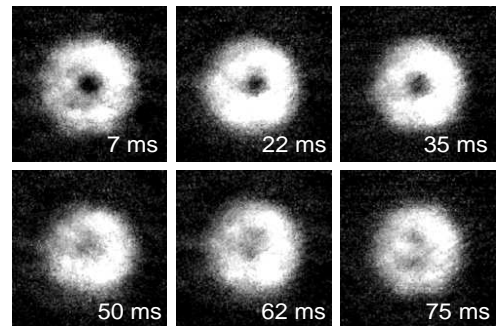}
\caption{Decay of a doubly quantized vortex. Axial absorption
images of condensates after 15~ms of ballistic expansion with a
variable hold time after imprinting a doubly quantized vortex. A
doubly quantized vortex decayed into two singly quantized
vortices. For this data, the interaction strength was
$an_{z}\approx 7.5$ (see text for definition). The field of view
in each image is 320~$\mu$m $\times$ 320~$\mu$m.
\label{f:evolution}}
\end{center}
\end{figure}

The decay of a doubly-quantized vortex state was studied by taking
an absorption images along the imprinted vortex line after
releasing the condensate and letting it expand for 15~ms. When we
took an integrated absorption image, the visibility of a vortex
core completely vanished within 30~ms. To reduce blurring due to
possible bending of the vortex line~\cite{RBD02}, we employed a
tomographic imaging technique~\cite{ATM97}. A 30~$\mu$m thick
central slice of the condensate (see Fig.~\ref{f:chip}(b)) was
selectively pumped into the $F=2$ hyperfine level with a sheet of
laser light perpendicular to the condensate long axis; the radial
profile of the condensate in the selected region was then imaged
with a light pulse resonant with the $F=2 \rightarrow F'=3$
cycling transition. In our absorption images, the size of a
doubly-quantized vortex core was typically $\sim 40~\mu$m. This
tomographic imaging technique was crucial for observing the time
evolution of vortex cores beyond 30~ms.

A series of absorption images of the splitting process is provided
in Fig.~\ref{f:evolution}. Images taken just after imprinting show
a doubly-quantized vortex core of high visibility; the visibility
of the core decreased with time, an effect we attribute to bending
of the vortex line~\cite{RBD02} and other excitations created
during the imprinting process. Later in the evolution, the central
core deformed into an elliptical shape and split into two
closely-spaced cores. Once the two cores were separated by their
diameter, they appeared well resolved in our images. The angular
position of the two cores was random for each experimental
realization with the same evolution time, so the precession
frequency of two cores could not be determined with our
destructive image technique.

\begin{figure}
\begin{center}
\includegraphics{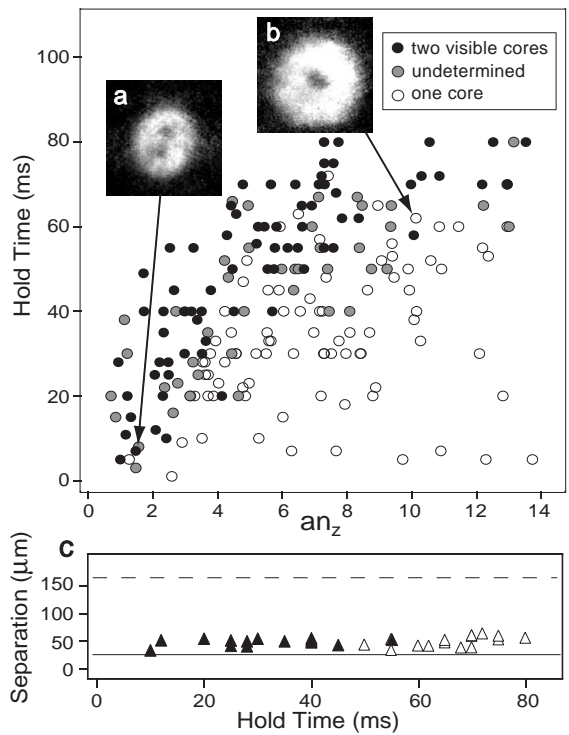}
\caption{Density dependence of the decay process. The time scale
for the decay process of doubly quantized vortex states was
measured by observing the vortex cores and classifying them as one
vortex (open circles) or two vortices (solid circles). Data were
collected with three axial trap frequencies $f_z=$~2.7, 3.7,
12.1~Hz and the interaction strength $an_z$ was controlled by
changing the atom number by rf induced evaporation before
imprinting. Typical absorption images for (a) fast decay at low
density ($an_z=1.5$) and (b) slow decay at high density
($an_z=10.1$). The field of view in the absorption images is
300~$\mu$m $\times$ 300~$\mu$m. (c) The separation of two visible
cores vs. the hold time for $2<an_z<3$ (solid triangles) and
$6<an_z<8$ (open triangles). The solid and dashed lines indicate
the diameter of one vortex core and of the condensate,
respectively.\label{f:density}}
\end{center}
\end{figure}

\begin{figure}
\begin{center}
\includegraphics{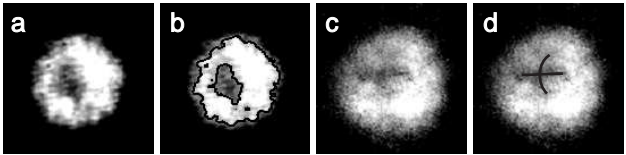}
\caption{Examples for the dynamic evolution after imprinting a
doubly quantized vortex: (a) Surface Exicitation. Regular density
modulation of the surface was observed after 51~ms hold time for
$an_z=1.8$ (b) same as (a) with a contour line. (c) Crossing of
vortex lines. 55~ms hold time and $an_z=8.4$. (d) same as (c) with
guide lines for vortex lines. The field of view is 270~$\mu$m
$\times$ 270~$\mu$m. \label{f:surface}}
\end{center}
\end{figure}

To investigate the dependence of the instability on the mean field
atomic interaction, we measured the characteristic time scale of
splitting of a doubly-quantized vortex core as a function of the
atom density. Atom density was controlled by removing a variable
number of atoms with rf evaporation before imprinting a vortex.
Images were classified as follows: images where the two cores were
separated by more than one core diameter were labelled as ``two
visible cores''; images with a clearly-defined circular central
core were labelled as ``one core''; images in the intermediate
range, where the central core was elliptical but the two cores
were not resolved, or with a bad visibility were labelled as
``undetermined''. For example, the images at 62~ms and 75~ms in
Fig.~\ref{f:evolution} and Fig.~\ref{f:density}(a) were classified
as ``two visible cores'', and 50~ms in Fig.~\ref{f:evolution}, and
Fig.~\ref{f:surface}(a) and (c) as ``undetermined''.

Experimental results are provided in Fig.~\ref{f:density} as a
function of the linear atom density $n_z$ (along the condensate
long axis) multiplied by the $s$-wave scattering length $a$. The
rescaled quantity, $an_z=a\int |\psi(r)|^2 dxdy$ corresponds for a
cylindrical condensate to the strength of the mean field
interaction, with $\psi(r)$ being the condensate wavefunction.
Results in Fig.~\ref{f:density} clearly demonstrate that a
doubly-quantized vortex core splits more slowly as the density
becomes higher.

Once the doubly-quantized vortex core split into two cores, the
distance between the two cores was almost constant ($\sim
50~\mu$m) during the further evolution, as shown in
Fig.~\ref{f:density}(c). This is evidence that the separation
process was driven mainly by the dynamical instability, and not by
dissipation, which would gradually increase the separation of the
two cores. Dissipative processes were minimized by performing the
experiments at the lowest possible temperature. Condensates did
not have any discernible thermal atoms even after extended hold
time. Furthermore, the energy released by the dissociation of the
doubly-quantized vortex was $\sim 5$~nK negligible to the critical
temperature $\sim 240$~nK. For the upper bound to the temperature
of $<100~$nK, Ref.~\cite{FS99} predicts that dissipative decay
time to be $\approx 1.5$~s for a single vortex, a time scale much
longer than what we observed.

Multiply-quantized vortices in a harmonic potential are predicted
to spontaneously decay into other states even in the absence of
dissipation and external perturbations~\cite{PLE99}. In the
Bogoliubov framework, which is believed to well describe quantized
vortices in one component condensates, the dynamical instability
manifests as the existence of excitation modes with a complex
eigenfrequency. The nonvanishing imaginary part of the
eigenfrequency implies an exponential growth in time of the
corresponding excitation mode, leading to decay of the
multiply-quantized vortex state. This spectral instability is a
general parametric phenomenon occurring when several modes compete
during coherent evolution and has been studied in many other
nonlinear physical systems (see, {\it e.g.},
Ref.~\cite{Ang03,ABA89} and references therein).

For a doubly-quantized vortex state in a cylindrically symmetric
condensate, it was theoretically found that there are two
excitation modes with a complex eigenfrequency~\cite{PLE99,
MMI03}. One of them is confined inside the doubly-quantized vortex
core; the growth of this so-called ``core'' mode induces splitting
of the original doubly-quantized vortex core into two separate
singly-quantized vortex cores. The other mode, having the
conjugate eigenfrequency, grows with the core mode in order to
conserve energy. In the low density limit, this mode corresponds
to the co-rotating quadrupole mode, leading to oscillations in the
surface shape of condensates. We always observed that the surface
of condensates changed into a quadrupole shape as the two cores
appeared, as shown in Fig.~\ref{f:density}(a), and the ellipticity
was larger at lower density.

The dynamical instability of the doubly-quantized vortex state is
related to the magnitude of the imaginary part of the complex
eigenfrequency, and, according to the numeric calculation in
Ref.~\cite{MMI03}, nonvanishing imaginary part of the
eigenfrequency appears at $an_z <3$ and $an_z \sim 12$, showing a
quasi-periodic behavior as a function of the interaction strength,
$an_z$. The experiment showed a monotonic increase of the lifetime
with no hint of periodic behavior. However, the calculated
instability is not directly comparable to the observed lifetime.
The imaginary part represents only the initial instability. Our
criterion for decay was the observation of two \emph{separated}
vortex cores. It is possible that the dynamical instability
changes after the doubly-quantized vortex state is significantly
perturbed~\cite{GP99,IHS03}. It would be helpful to have more
inclusive calculations leading to a lifetime directly comparable
with the experiments.

What is the further evolution of the two cores? Some of the images
at low density showed a regular surface modulation, as in
Fig.~\ref{f:surface}(a), which was not seen in clouds with a
single core. This indicates that higher-order surface modes are
excited during the coherent evolution~\cite{KTU03}. However, their
reproducibility was insufficient for a systematic study. Several
images, especially those labelled as ``undetermined'', suggest
that vortex lines crossed~\cite{MMI03,GP01}, as in
Fig.~\ref{f:surface}(c). In our system, it was difficult to trace
the positions of the two cores beyond 80~ms hold time.

In conclusion, we observed how a doubly-quantized vortex splits
into a pair of singly-quantized vortices, and found higher
stability at higher atom density. The topological phase imprinting
technique is unique in generating doubly- or quadruply-quantized
vortex states~\cite{LGC02,KO04}; a key feature is the rapid
preparation of well-determined vortex states which gives access to
their dynamical instabilities and coherent evolution.

This work was funded by ARO, NSF, ONR, and NASA. M.S. acknowledges
additional support from the Swiss National Science Foundation and
C.S. from the Studienstiftung des deutschen Volkes. We thank M.
M\"{o}tt\"{o}nen and K. Machida for helpful discussions.

\end{document}